# Reduced Area Low Power High Throughput BCD Adders for IEEE 754r Format

Himanshu Thapliyal
Centre for VLSI Design
IIIT Hyderabad, India
(thapliyalhimanshu@yahoo.com)

Hamid R. Arabnia
Department of Computer Science
The University of Georgia, U.S.A
(hra@cs.uga.edu)

M.B Srinivas
Centre for VLSI Design
IIIT Hyderabad, India
(srinivas@iiit.net)

## ABSTRACT

IEEE 754r is the ongoing revision to the IEEE 754 floating point standard and a major enhancement to the standard is the addition of decimal format. Firstly, this paper proposes novel two transistor AND and OR gates. The proposed AND gate has no power supply, thus it can be referred as the Powerless AND gate. Similarly, the proposed two transistor OR gate has no ground and can be referred as Groundless OR. Secondly for IEEE 754r format, two novel BCD adders called carry skip and carry look-ahead BCD adders are also proposed in this paper. In order to design the carry look-ahead BCD adder, a novel 4 bit carry look-ahead adder called NCLA is proposed which forms the basic building block of the proposed carry look-ahead BCD adder. Finally, the proposed two transistors AND and OR gates are used to provide the optimized small area low power high throughput circuitries of the proposed BCD adders.

## 1. INTRODUCTION

Nowadays, the decimal arithmetic is receiving significant attention as the financial, commercial, and Internet-based applications cannot tolerate errors generated by conversion between decimal and binary formats. Furthermore, a number of decimal numbers, such as 0.110, cannot be exactly represented in binary, thus, these applications often store data in decimal format and process data using decimal arithmetic software [1]. The advantage of decimal arithmetic in eliminating conversion errors also comes with a drawback; it is typically 100 to 1,000 times slower than binary arithmetic implemented in hardware.

Since, the decimal arithmetic is getting significant attention; specifications for it have recently been added to the draft revision of the IEEE 754 standard for Floating-Point Arithmetic. IEEE 754r is an ongoing revision to the IEEE 754 floating point standard [2,3]. Some of the major enhancements so far incorporated are the addition of 128-bit and decimal formats. Furthermore, three new decimal formats are described, matching the lengths of the binary formats. These have led to the decimal formats with 7, 16, and 34-digit significands, which may be normalized or unnormalized. In the proposed IEEE 754r format, for maximum range and precision, the formats merge part of the exponent and significand into a combination field, and compress the remainder of the significand using densely packed decimal encoding [2,3]. It is anticipated that once the IEEE 754r Standard is finally approved, hardware support for decimal floating-point arithmetic on the processors will come into existence for financial, commercial, and Internet-based applications. Still, the major consideration while implementing BCD arithmetic will be to enhance its speed as much as possible.

Furthermore, in order to satisfy the Moore's law and the high speed processing needs, more and more logic elements are packed into smaller and smaller volumes and are clocked at higher and higher frequencies, dissipating more and more heat. This huge dissipation of heat leads to exhaustion of batteries of portable systems and systems overheat [4]. Thus, today there is an increasing need of the portable applications requiring small-area low-power high throughput circuitry.

In order to satisfy all the aforesaid expectations; firstly, this paper introduces novel two transistors AND and OR gates. The proposed two transistor AND and OR gates have only either VDD or GND, thus preventing the flow of short circuit current. Secondly, two novel BCD adder architectures termed CLA BCD (Carry look Ahead BCD) and CS BCD( Carry Skip BCD) adders are also proposed in this paper. In order to propose the novel CLA BCD adder, a new 4-bit carry look-ahead architecture called NCLA is proposed which is better than the fastest CLA [7], in terms of area while maintaining the speed improvement of the fastest CLA. The carry skip BCD adder is proposed to cater the need the need of carry skip adder in decimal arithmetic. Finally, the proposed two transistors AND and OR gates are used to provide the optimize implementation of the proposed BCD adders in terms of small area, low power and high throughput.

## 2. TWO TRANSISTOR AND/OR GATES

In this paper, novel two transistor AND and OR gates are proposed. The proposed two transistors AND gate is shown in Figure 1. It has no power supply, thus it can be referred as the Powerless AND gate. Similarly, two transistor OR gate is proposed as shown in Figure 2, having no ground and can be referred as Groundless OR gate. As far as the survey of literature and our knowledge is concerned, these are the most optimal designs of the AND and OR gates which are directly scalable to higher inputs by cascading.

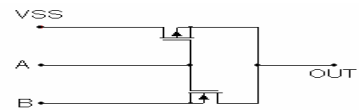

Figure 1. Proposed Two Transistor AND Gate

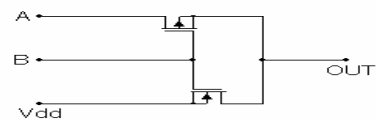

Figure 2. Proposed Two Transistor OR Gate

The evaluation of the proposed architectures is done both theoretically as well as experimentally through SPICE simulations.

The following equation [5] is used to estimate the power consumption of a circuit.

$$P_{Dynamic} = (\sum_i C_i * V_{i\_swing} * P_i) * f_{clk} + i_{sc} * VDD + \sum_i i_{leak} * VDD$$

In the above equation, $P_{Dynamic}$ is the dynamic power consumption, $C_i$ is the load capacitance, $V_i$ is the voltage swing, $P_i$ is the probability of a switch, $f_{clk}$ is the clock frequency, $I_{sc}$ is the short-circuit current, $I_{leak}$ is the leakage current and $VDD$ is the supply voltage. It can be inferred from the above equation that the main components of the power dissipation are the $I_{sc}$ and $V_{swing}$ components. Since, the $I_{leak}$ component in the equation is usually very low and is generally discarded. The voltage swing of a circuit is referred as the change in voltage during a transition and is equal to the voltage difference between logic '1' and logic '0'. When the signal is perfectly transmitted, the logic '1' is equal to VDD and the logic '0' is equal to VSS. Hence, the voltage swing is equal to the supply voltage and a reduction in the supply voltage will result in lower power dissipation. There is a reduction in the voltage swing when the signals are not fully transmitted which is generally occurred when an NMOS transmits logic '1' or a PMOS transmits logic '0'. Moreover, the circuit that has a lesser driving capability often dissipates less power [6].

The proposed AND and OR gate has incomplete voltage swings since in the proposed AND gate PMOS is permanently tied to logic '0' and in the proposed OR gate NMOS is permanently tied to logic '1'. Moreover, since the proposed gates have only either VDD or GND, neither both in the same circuit, there is no short-circuit current established by a direct path between VDD and VSS. Thus the proposed gates are highly optimized in terms of power consumption.

## 3. EXPERIMENT DESCRIPTION

The simulation environment is setup to measure the performance of the proposed circuits in terms of propagation delay and power dissipation. The simulations are performed by varying the frequencies and the capacitive loads to ensure that the proposed circuits work at different frequencies and different capacitive loads. The Simulation conditions are shown in Figure 3.

| Load | 0.01 pf | 0.02pf | 0.05pf | 0.1pf | 0.3pf | 0.5pf |
|---|---|---|---|---|---|---|
| Frequencies | 1 Mhz | 50 Mhz | 100 Mhz | 200Mhz | | |

Figure 3. Simulation Conditions

The simulation is done in Tanner spice and the technology being used is 0.35-um CMOS digital technology (TSMC 35, Canadian Microelectronic Corporation) with a 3.3-V supply voltage. The propagation delay is measured when the changing input reaches 50% of the transition to the time when the output reaches its 50%. Figure 4 and Figure 5 shows the power consumption and the propagation delay for a load of 0.01pf at different frequencies. Similarly results have been obtained for the other loading conditions. Since the proposed gates are implemented with the bare minimum of two transistors and are most optimal, hence no comparative study is shown.

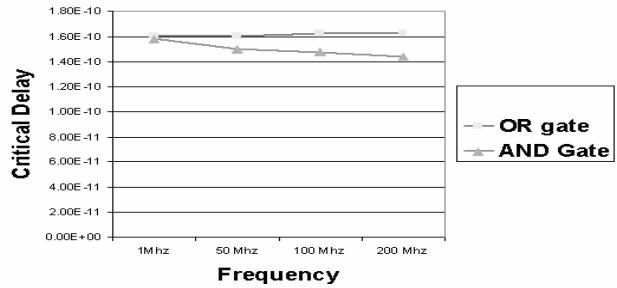

Figure 4. Critical Delay of Proposed Circuits(Seconds) at 0.01pf

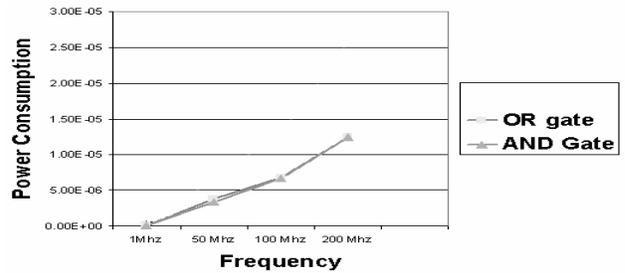

Figure 5. Power Consumption of Proposed Circuits(Watts) at 0.01pf

## 4. PROPOSED NEW CARRY LOOK- AHEAD (NCLA) ADDER

Recently, a modified carry look-ahead adder [7] (abbreviated as MCLA) is proposed which is similar to CLA (carry look-ahead adder) in basic construction. The MCLA circuit in [7] uses NAND gates to replace the AND, OR and NOT gates in the conventional CLA to decrease the cost and increase the speed. It can be easily inferred from the design of MCLA [7] that in spite of having significant speed improvement, it has significant increase in area due to excessive number of NAND gates used for the faster carry propagation. This problem will significantly increase when the MCLA will be cascaded for designing higher order CLA. Thus, the authors propose a new CLA architecture called NCLA emphasizing the use of AND and OR gates instead of NAND gates. It has already been proved above that the proposed AND and OR gates are highly optimized in terms of area, low power and speed. Thus implementing the AND and OR functions in the conventional CLA with the proposed AND and OR gates will significantly increase its performance compared to MCLA using NAND gates, in terms of area, power and throughput. The proposed NCLA will consist of PGA(propagate generate architecture) block that is used to generate Si, Gi and Pi as shown in Figure 6. The proposed 4-bit NCLA adder is shown in Figure 7, in which at the 4th place full adder is used instead of PGA. This is done to reduce the area (number of gates) of the NCLA without sacrificing the speed improvement found in MCLA. It can be easily be verified that there will be reduction of number of gates to generate the final carry as shown in the Figure 7. Using the proposed AND and OR gates, the number of transistors required to implement the proposed NCLA are 74 compared to the use of 136 transistors in MCLA proposed in [7]. The Comparison is clearly explained below. Table I shows the number of transistors required to implement the various logic operations.

TABLE I. NUMBER OF TRANSISTORS REQUIRED TO PERFORM DESIGN THE LOGIC OPERATIONS

| Logic Function | Number of Transistors Required |
|---|---|
| NOT Gate | 2 |
| 2 input XOR gate | 4 |
| 2 input NAND gate | 4 |
| 3 input NAND gate | 6 |
| 4 input NAND gate | 8 |
| 1 bit full adder[8] | 10 |
| 1 bit multiplexer based full adder without any Power supply[4] | 12 |
| Proposed 2 input AND gate | 2 |
| Proposed 2 input OR gate | 2 |
| 3 input AND gate designed by cascading proposed AND gate | 4 |
| 4 input AND gate designed by cascading the proposed AND gate | 6 |
| 4 input OR gate designed by cascading the proposed AND gate | 6 |

The metamorphosis of partial full adder(MPFA) in [7] generating ($S_i, G_i, P_i$) requires two two-input XOR gate and one two-input NAND gate leading to the total number of 12 transistors as shown below

2   2_Ex-Or

1   2_NAND

=> 2*(4) + 1*(4) = 12 transistors.

Table II shows the total number of transistors required to implement the fastest CLA called MCLA is 136.

TABLE II. NUMBER OF TRANSISTORS REQUIRED TO IMPLEMENT MCLA[7]

| Blocks | Number of Transistors |
|---|---|
| 4  MPFA | 12 * 4= 48 |
| 3   NOT | 3 *2=6 |
| 5  two input NAND | 5*4=20 |
| 4   three input NAND | 4*6=24 |
| 4    four input NAND | 4 *8=32 |
| 1   four input AND | 6 |
| Total Number of Transistor | 136 |

The proposed CLA in this paper requires 74 transistors as shown in the Table III.

TABLE III. NUMBER OF TRANSISTORS REQUIRED TO IMPLEMENT NCLA

| Gates | Number of Transistors |
|---|---|
| 3  1 bit NFA | 3 * 10=30 |
| 3  two input AND gate | 3 *2=6 |
| 2  three input AND gate | 2 *4=8 |
| 1  four input AND gate | 6 |
| 1  two input OR gate | 2 |
| 1  three input OR gate | 4 |
| 1  four input OR gate | 6 |
| 1 bit Adder | 12 |
| **Total number of Transistors** | **74** |

Thus, it can be easily inferred from Table II and Table III that the proposed NCLA is highly beneficial in number of transistors, thus in terms of area, speed and power compared to MCLA proposed in [7]. Figure 8 shows the transistor implementation of the PGA in which XOR gate proposed in [9] is used and Figure 9 shows the transistor implementation of the proposed NCLA in which the fourth block full adder is the multiplexer based low power adder proposed in [4]. The multiplexer based adder is used to further improve the speed and have the more power saving in NCLA. The proposed 4-bit NCLA is used to design optimized CLA BCD adder as explained in the next section. The functional verification of the NCLA is done using Verilog HDL at gate level and using SPICE at transistor level.

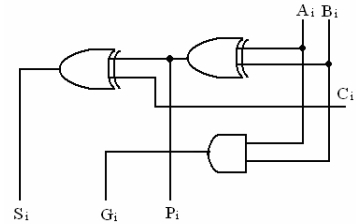

Figure 6.  PGA Block generating $S_i$, $G_i$ and $P_i$

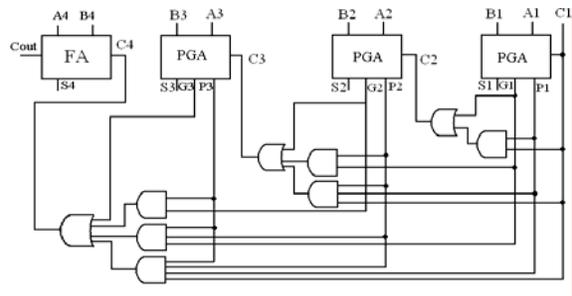

Figure 7.  Proposed 4-bit NCLA

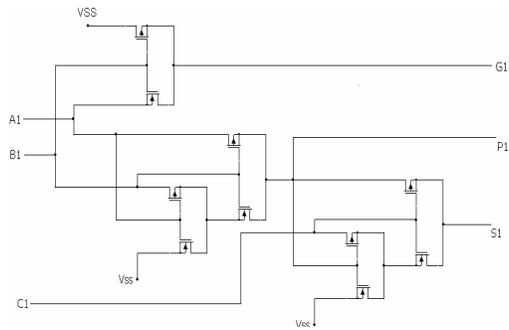

Figure 8. Transistor implementation of PGA

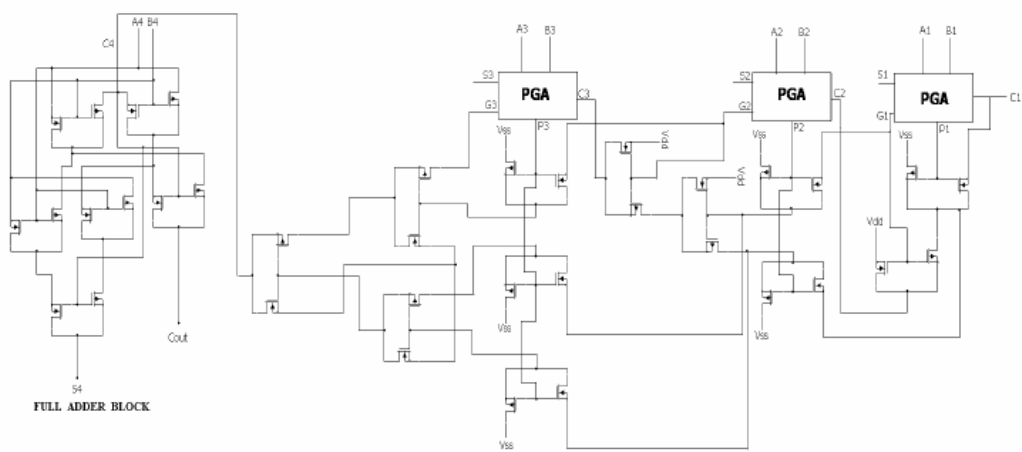

Figure 9. Transistor Implementaion of NCLA using proposed AND and OR gates and Multiplexer based full adder

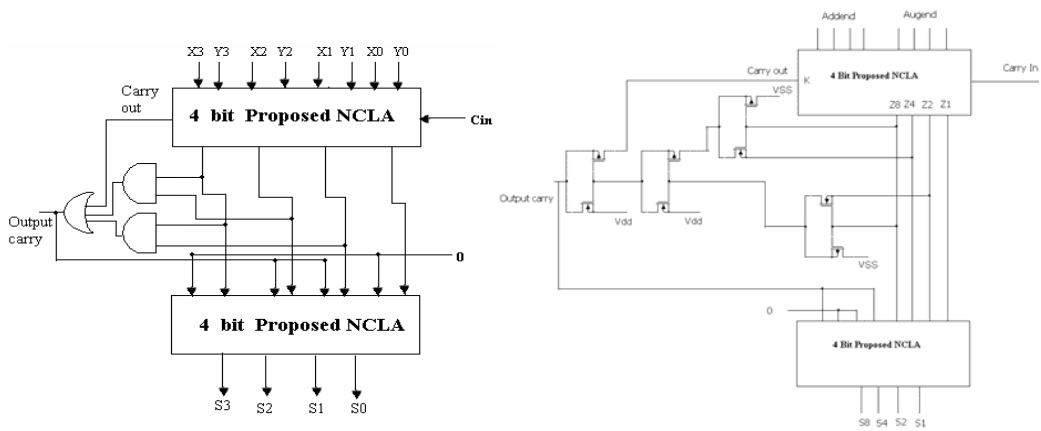

Figure 10. (a) Proposed CLA BCD Adder (b) Trannsitor implementation of the proposed CLA BCD adder

## 5. PROPOSED CARRY LOOK-AHEAD BCD ADDER

In this paper, the authors also propose a novel Carry Look-Ahead BCD Adder which is suitable for CMOS implementation. In the proposed CLA BCD adder, the 4-bit adders used in the conventional BCD adder are replaced by the proposed 4-bit NCLA. The two decimal digits, together with the input carry, are first added in the top 4-bit NCLA to produce the binary sum. When the output carry is equal to zero, nothing is added to the binary sum. When it is equal to one, binary 0110 is added to the binary sum using another 4-bit NCLA (bottom NCLA). The output carry generated from the bottom NCLA is ignored, since it supplies information already available at the output carry terminal. Furthermore, the OR and AND function used in the conventional BCD adder are implemented with the proposed AND and OR gates. Figure 10(a) shows the proposed CLA BCD adder and Figure 10(b) show the transistor implementation of the proposed CLA BCD adder using proposed AND and OR gates. The functional verification of the proposed CLA BCD adder is done in Verilog HDL at gate level using Model Sim and at transistor level using TSPICE at 0.35 micron TSMC library.

## 6. PROPOSED CARRY SKIP BCD ADDER

The proposed Carry Skip BCD Adder is being constructed in such a way that, the first full adder block consisting of 4 full adders can generate the output carry 'Cout' instantaneously, depending on the input signal and 'Cin', without waiting for the carry to be propagated in the ripple carry fashion. Figure 11 shows the proposed Carry Skip BCD adder. The working of the proposed Carry Skip BCD Adder (CS BCD Adder) can be explained as follows.

In the single bit full adder operation, if either input is a logical one, the cell will propagate the carry input to its carry output. Hence, the ith full adder carry input $C_i$, will propagate to its carry output, $C_{i+1}$, when $P_i = X_i \oplus Y_i$ where $X_i$ and $Y_i$ represents the input signal to the ith full adder. In addition, the four full adders at the first level making a block can generate a "block" propagate signal 'P'. When 'P' is one, it will make the block carry input 'Cin', to propagate as the carry output 'Cout' of the BCD adder, without waiting for the actual propagation of carry, in the ripple carry fashion. An AND gate is used to generate a block propagate signal 'P'. Furthermore, depending on the value of 'Cout', appropriate action is taken. When it is equal to one, binary 0110 is added to the binary sum using another 4-bit binary adder (Second level or bottom binary adder). The output carry generated from the bottom binary adder is ignored, since it supplies information already available at the output carry terminal. Figure 12 shows the transistor level implementation of the proposed carry skip BCD adder in which the 10T (10 Transistor) full adder proposed in [8] and the XOR proposed in [9] are used for implementing the full adder and the XOR gate respectively. The proposed AND and OR gates are used for generating the OR and AND functions. The functional verification of the proposed carry skip BCD adder is done in Verilog HDL at gate level while at transistor level; the functionality is verified using TSPICE 0.35 micron technology.

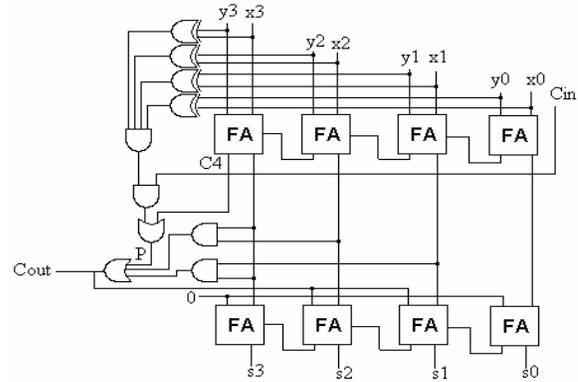

Figure 11. Proposed Carry Skip BCD Adder

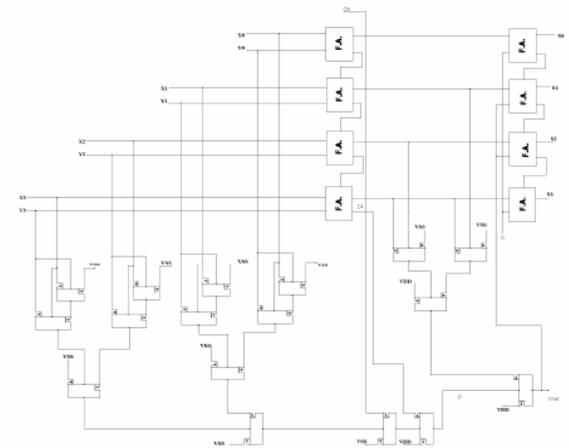

Figure 12. Transistor Level Implementaion of the Proposed Carry Skip BCD Adder

## 7. CONCLUSIONS

The focus of this paper is the IEEE 754r which is the ongoing revision to the IEEE 754 floating point standard considering decimal arithmetic. Thus, this paper proposes two transistors AND and OR gates for small area low power high throughput circuitries. Two novel designs of BCD adder called Carry Look Ahead and Carry Skip BCD Adders are also proposed in this paper to cater the need of IEEE 754r format. A novel 4-bit carry look-ahead adder called NCLA is proposed which is used to design the carry look-ahead BCD adder. Finally the optimize transistor implementation of the proposed BCD adders is presented using the proposed AND and OR gates. The circuits are designed in such a manner that they are highly optimized in terms of small-area low-power and high throughput. The authors believe that the proposed designs will provide a platform for designing high speed low power digital circuits such as Arithmetic and logical unit implemented in digital signal processors.